\documentclass[12pt,BCOR2mm,DIV12]{scrartcl}

\usepackage{graphicx}
\usepackage{amssymb,amsfonts,amsmath,mathtools,array}
\usepackage{booktabs}
\usepackage[british]{babel}
\usepackage{microtype}
\usepackage{algorithm}

\graphicspath{{graphics/}}
\usepackage{bm}
\usepackage{hyperref}
\usepackage{xcolor}
\hypersetup{
    colorlinks,
    linkcolor={red!50!black},
    citecolor={blue!50!black},
    urlcolor={blue!80!black}
}
\usepackage{natbib}
\begin{document}
\title{Diagnostics for assessing the linear noise and moment
closure approximations} 
\author{Colin S. Gillespie and Andrew Golightly}
\maketitle

\begin{abstract}
  Solving the chemical master equation exactly is typically not possible, so
  instead we must rely on simulation based methods. Unfortunately, drawing exact
  realisations, results in simulating every reaction that occurs. This will
  preclude the use of exact simulators for models of any realistic size and so
  approximate algorithms become important. In this paper we describe a general
  framework for assessing the accuracy of the linear noise and two moment
  approximations. By constructing an efficient space filling design over the
  parameter region of interest, we present a number of useful diagnostic tools
  that aids modellers in assessing whether the approximation is suitable. In
  particular, we leverage the normality assumption of the linear noise and
  moment closure approximations.
 
\end{abstract}

\section{Introduction} \label{s:intro}

Due to advances in experimental techniques, it is now clear that cellular
dynamics incorporate a vast array of heterogeneous components. Whilst each
component may be relatively simple, combining component systems results in
complex, temporal dynamics that are not amenable to simple intuitive
understanding.

The recognition of such biological sophistication has lead to the conclusion
that complex biological processes cannot be understood through the application
of ever-more reductionist experimental programs. Instead, by formulating the
system of interest into a mathematical framework, we can begin to combine
disparate sources of knowledge. Furthermore, careful mathematical modelling of
biological processes has other advantages. For example, \cite{Kowald1996}
highlight possible interactions that would be difficult to observe
experimentally. Therefore, a successful analysis of a biological system now
requires a complementary wet and dry approach (see \cite{Ingalls2008}).

When modelling biological networks, it is important to incorporate the intrinsic
noise of the system. One standard approach is to utilise stochastic kinetic
models described using a set of chemical reactions, their associated hazards 
and an assumption that the system evolves according to a continuous-time 
Markov jump process (MJP). The transition kernel governing the MJP can be found 
by constructing and solving Kolmogorov's forward equation, known in this context 
as the chemical master equation (CME) \citep{Gillespie1992}. Unfortunately, the 
CME is rarely tractable for systems of interest and the vast
size of the underlying state space means that numerically computing the solution
of the CME is not feasible (see \cite{Wilkinson:2012}). While it may not be
possible to solve the chemical master equation, it is usually straightforward to
obtain exact realisations of the MJP using standard simulation algorithms. The most well known
algorithm is the \textit{direct method} developed by \cite{Gillespie1976}.

Simulating from the model is not only crucial when building a system, but is
also essential for parameter inference, since the observed data likelihood is
usually analytically intractable. Exact simulation based approaches to MJP inference 
typically use data augmentation \citep{Boys:2008} coupled with Markov chain 
Monte Carlo (MCMC) or particle MCMC \citep{Golightly:2011,Owen:2015}. In the simplest 
implementation of the latter, only forward simulations from the model are 
required, and the method can be regarded as likelihood-free. Other likelihood-free 
approaches include the use of approximate Bayesian computation (ABC) schemes 
\citep[see for example,][]{Beaumont:2002,Sisson:2007,Toni2009}. These inference 
schemes typically require many millions of forward simulations and the resulting 
computational cost may preclude their use when the system size or reaction rates 
are large. Due to this computational hurdle, a number of approximate simulators 
have been proposed \citep[for an overview,see][]{Pahle2009}. Use of an approximate 
simulator in this way can be seen as performing exact (simulation-based) inference 
for the associated approximate model. 


Approximate models (and their associated simulators) that ignore the discrete 
nature of the stochastic kinetic model, but crucially, not stochasticity, include the diffusion 
approximation \citep{Gillespie:2000}, the linear noise approximation (LNA) \citep{Kurtz:1970,Elf:2003} and moment closure approaches 
\citep{Kampen2007,Gillespie2009a}). Hybrid approaches which treat some species 
as discrete and others as continuous have been proposed by \cite{Salis2005} 
and \cite{Sherlock:2014} among others. Moment closure and LNA based approaches 
are particularly attractive, due to their tractability. For the former, the first two 
moments of the MJP are combined with an assumption of normality, whereas for the latter, 
the CME is approximated in a linear way, to give a process with normal transition 
densities. Unfortunately it is not straightforward to check whether a given approximation
technique yields acceptable results since, by definition, the approximate
simulator is not exact. For example, if the model contains \textbf{any}
second-order reactions then the mean population estimate from the linear noise
approximation will not be exact (see \cite{Golightly2013b} for example).
However, the approximation may still be sufficient for model exploration or
parameter inference.


Recently, \cite{Cao2006} and \cite{Jenkinson2013} performed a comparison of approximate and
exact simulators at specific parameter values. Essentially, each proposed
simulating $N$ times from an exact and an approximate algorithm, calculating a
distance metric and assessing accuracy by performing a hypothesis test. However,
there are two major drawbacks with this test driven approach. First, for many
approximate simulators we can analytically prove that the approximate and exact
differ, so as $N$ increases we will \textit{always} reject the null hypothesis.
Second, in the parameter inference setting we are interested in the performance
of the approximate simulator across a range of parameter values, not just at a
particular value.

In the related field of computer experiments, complex models are emulated using
a faster model; typically a Gaussian process (GP). Since prediction is made
using an emulator, it is essential that the emulator accurately represents the
system. \cite{Bastos2009c} provide a number of useful diagnostic measures (in
the context of Gaussian processes) for assessing simulator quality. Within the 
context of stochastic kinetic models, both the moment closure and LNA approaches 
can be seen as GP emulators. 

In this paper we present a set of general, principled methods for efficiently
assessing the quality of the linear noise and moment closure approximations
across a large parameter space based on the techniques found in the computer
experiment literature. The diagnostic measures we present are simple to calculate 
and interpret, providing the practitioner with a useful tool for assessing 
simulator accuracy. The remainder of this paper is organised as follows. Section~2 
briefly reviews stochastic kinetic models and exact simulation techniques before 
introducing the moment closure and linear noise approximations. In Section~3 
we describe efficient methods for exploring the parameter
space, and the diagnostic measures which comparisons between simulators 
are to be based on. The methods are illustrated using three examples of increasing complexity.


\section{Stochastic kinetic models}

Suppose we have a system of chemical reactions with $u$ chemical species
$\{X_1,\ldots, X_u\}$ and $v$ reactions $\{R_1, \ldots, R_v\}$, where reaction
$R_k$, with rate parameter $c_k$, corresponds to
\[
\underline{s}_{k1} X_1+\ldots + \underline{s}_{ku} X_u
\rightarrow \bar{s}_{k1} X_1 + \ldots + \bar{s}_{ku}
X_u,
\]
with $\underline{s}_{ki}$ and $\bar{s}_{ki}$ the number of molecules of type
$X_i$ before and after the reaction $R_k$, respectively. Let $X_{j,t}$ be the
random variable denoting the number of molecules of species $X_j$ at time $t$
and let $X_t$ be the $u$-vector $X_t = (X_{1,t}, X_{2, t}, \ldots, X_{u, t})' $.
Further, let $\underline{s} = (\underline{s}_{ij})$ be a $v\times u$ matrix of
the coefficients $\underline{s}_{ij}$ with $\bar{s}$ being defined similarly.
Then the $u \times v$ stoichiometry matrix $s$ is defined by
\begin{equation}\label{1}
s = (\bar{s} -  \underline{s})' \;.
\end{equation}
We denote $x_{i,t}$ to be the number of molecules of species $X_{i}$ at time
$t$, and let $x_t$ be the $u$-vector $x_t = (x_{1,t}, x_{2,t}, \ldots, x_{u,t})'
$.

The rate of reaction $R_k$ is defined by the rate function $h_k(x_t, c_k)$,
where $c_k$ is the reaction rate constant. Hence, the hazard of a type $k$
reaction occurring depends on the rate constant $c_k$, as well as the state of
the system at time $t$. This system can be naturally modelled as a Markov jump
process, that is, in a small time increment, $\delta t$, the probability of
reaction $R_k$ occurring in the time interval $(t, t+ \delta t]$ is $h_k(x_t,
c_i)\delta t$ \citep{Gillespie1992}. When a reaction of type $k$ does occur, the system state changes
by $\bar{s}_{k}-\underline{s}_k$. A typical model assumption is that the reactions follow mass action kinetics.
This results in a hazard function that takes the form of the rate constant $c_k$
multiplied by a product of binomial coefficients expressing the number of ways
in which the reaction can occur.

The transition kernel of the MJP can be found by constructing and solving 
Kolmogorov's forward equation, known in this context as the chemical master equation
(CME). Denote $p(x_t)$ as
the probability of being in state $x_t$ and note that we suppress dependence 
of $p(x_t)$ on the initial state $x_0$ and the reaction constants $c = (c_1, \ldots, c_v)'$ 
for simplicity. The CME is given by
\begin{equation}\label{2}
  \frac{d}{dt} p(x_t) = \sum_{k=1}^v p(x_{t} - s_k)
  h_k(x_t - s_k, c_k) - p(x_t) h_k(x_t, c_k),
\end{equation}
where $h_k(x_t, c_k)$ is the hazard function for reaction $R_k$ and $s_k$ is the
$k^{\text{th}}$ column of the matrix $s$. Once $p(x_t)$ is obtained, a complete 
characterisation of the system is available. Unfortunately, 
the CME is only tractable for a handful of cases \citep[see e.g.][]{Gardiner1985}. 
Consequently, for most systems of interest, an analysis via the CME will not be possible.

\subsection{Exact simulation}

Although the chemical master equation is rarely analytically tractable, it is
straightforward to draw exact realisations using a discrete event simulation
method. The standard algorithm, developed by \cite{Gillespie1976}, for
simulating from a stochastic system is the direct method (described in Algorithm
1). Essentially, at each algorithm iteration we select a reaction to occur and
update the species levels and clock. However, as the number of reactions or the
size of the hazard functions increase, the computational cost increases.

A number of improvements to this algorithm have been proposed. For example,
\cite{McCollum2006} dynamically reorder the reactions from most to least likely,
to significantly increase the speed of the algorithm. Alternatively
\cite{Cao2004} suggest an pilot simulation to optimise the reaction order.
\cite{Gibson2000} exploit the model structure to avoid unnecessary updates.
However, the underlying speed issues still remain for models of reasonable size, 
necessitating the use of approximate simulation strategies. 

\begin{table}[!t]
  \centering
  \begin{tabular}{@{}ll@{}}
    \hline
    \multicolumn{2}{@{\,}l @{}}{\textbf{Algorithm 1:} Direct method \citep{Gillespie1976}}\\
    \hline
    {\small 1:} & Set $t=0$ and initialise rate constants $c_1, \ldots, c_v$ and
    the initial molecule \\
    & \quad numbers $x_{1,0}, \ldots, x_{u,0}$. \\
    {\small 2:} & Propensities update: update each of the $v$ hazard functions,
    $h_k(x_t, c_k)$ \\
    & \quad based on the current state, $x_t$.\\
    {\small 3:} & Calculate the total hazard, $h_0(x_t,c) = \sum_{k=1}^v h_k(x_t, c_k).$\\
    {\small 4:} &  Simulate the time to the next event, $\tau \sim Exp(h_0(x_t,
    c))$ and set $t = t + \tau$.\\
    {\small 5:} & Simulate the reaction index, $j$, with probabilities  $h_k(x_t,
    c_k)/h_0(x_t, c), i=1, \ldots, v$.\\
    {\small 6:} &  Update $x_t$ according to reaction $j$.\\
    {\small 7:} & If simulation time is exceeded, stop, otherwise return to step
    2. \\
    \hline
  \end{tabular}
\end{table}

\subsection{Normal approximations}\label{norm}

In what follows, we consider two tractable approximations of $p(x_t)$ 
that ignore discreteness but not stochasticity. Both approaches 
assume that the distribution of $X_t$ at a particular time point,
$t$, is normal, so that
\begin{equation} \label{gauss}
X_t \sim N(\psi_t(c), \Sigma_t(c))
\end{equation}
where we let the approximate mean and variance $\mathbf{\psi}_t(c) 
= \hat E(X_t)$ and $\Sigma_t(c) = \widehat Var(X_t)$ depend explicitly 
on the rate constants $c$. 
Thus the approximate density at a particular time point is 
\[
  \hat p(x_t) = \frac{1}{(2 \pi)^{u/2} \vert
    \Sigma_t(c)\vert^{1/2}} 
  \exp\left[\frac{1}{2} (x_t -\psi_t(c))'\,
    [\Sigma_t(c)]^{-1} (x_t - \psi_t(c))\right] \;.
\]
It remains that we can choose appropriate forms for $\psi_t(c)$ and 
$\Sigma_t(c)$. We consider two related approaches, namely, 
moment closure and the linear noise approximation (LNA). We give a brief, 
informal description of these techniques in the sequel, and refer 
the reader to \cite{Kampen2007} and \cite{Wilkinson:2012} for further 
discussion.

\subsubsection{Moment closure}

Here, we approximate the moment equations
of the system as a set of ordinary differential equations (ODEs). These
equations then provide estimates of the mean and variances of individual
chemical species.

To extract the moment equations using the moment closure assumption we first
define the moment generating function (indexed by $\theta$) as
\begin{equation}\label{3}
M(\theta;t) = \sum_{x_t} p(x_t) e^{\theta x_t} \;.
\end{equation}
The moments, $E[x_t^n]$, where $x_t^n=(x_{1,t}^{n_1}, \ldots, x_{u,t}^{n_u})'$,
of the joint probability distribution can be found by taking $n^{th}$ order
derivatives of the moment generating function with respect to $\theta =
(\theta_1, \ldots, \theta_u)'$. The first moment is the mean and the second
moment can be used to obtain the variance.

On multiplying the chemical master equation (\ref{2}) by $e^{\theta x_t}$ and
summing over $x_t$ gives
\begin{equation}\label{4}
\frac{\partial M(\theta; t)}{\partial t} = 
\sum_{x_t} e^{\theta x_t} \sum_{k=1}^v p(x_t - s_k)
  h_k(x_t - s_k, c_k) - p(x_t) h_k(x_t, c_k)\;.
\end{equation}
The time evolution of the mean concentration of species $X_i$ can be obtained by
taking the first derivative of equation (\ref{4}) with respect to $\theta_i$ and
then setting $\theta$ to zero. Differentiating equation (\ref{4}) twice with
respect to $\theta_i$ yields $E[x_{i,t}^2]$, from which we can obtain the
variance. Similarly, differentiating with respect to $\theta_i \theta_j$ gives
$E[x_{i,t}\, x_{j,t}]$.

Following this process, we can obtain an ordinary differential equation (ODE)
for any moment of interest. However when we have non-linear dynamics, the
equation for the $i^{th}$ moment generally depends on the the $(i+1)^{th}$
moment equation, i.e. the ODE for the mean contains a term depending on the
second order moment. To circumvent this problem, we need to \textit{close} the
system, for example, by assuming an underlying Gaussian distribution. The mean and 
variance in (\ref{gauss}), which we denote by $\psi_t^m(c)$ and 
$\Sigma_t^m(c)$ in this context, are then easily obtained

\cite{Grima2012} \citep[see also][]{Singh2007,Smadbeck2013} 
shows that increased accuracy of lower-order moment estimation
can be obtained by using a higher-order closure scheme. However even though we
can estimate higher order moments, it is not clear how these estimates can be
routinely utilised. Hence a popular closure choice is to assume normality,
resulting in coupled equations for only the mean and variance. This particular
closure is also known as the two moment approximation (2MA).

\subsubsection{Linear noise approximation}

The linear noise approximation can be formed by first constructing the chemical
Langevin equation (CLE). In an infinitesimal time interval $(t, t + dt]$, the
reaction hazards will remain constant almost surely. This allows us to treat the
occurrence of reaction events as the occurrence of events from a Poisson process
with independent realisations for each reaction type. Writing $dR_t$ for the
$v$-vector number of reaction events of each type in the time increment, it then
follows that $\text{E}(dR_t) = h(X_t, c)dt$ and $\text{Var}(dR_t) = \text{diag}
\{h(X_t,c)\}dt$. Using $dX_t = s d R_t$ and matching 
$\text{E}(dX_t)$ and $\text{Var}(dX_t)$ with the drift and diffusion coefficients 
of an It\^o stochastic differential equation (SDE) gives
\begin{equation} \label{8}
d X_t = s h(X_t, c) dt + \sqrt{s \text{diag}\{ h(X_t, c)\} s'} dW_t 
\end{equation}
where $dW_t$ is the $u$-dimensional Brownian motion increment. 
Equation (\ref{8}) is commonly referred to as the chemical Langevin equation
(CLE). 

The LNA can now be derived from the CLE as follows. We replace 
the hazard function in equation (\ref{8}) with the rescaled form
$\Omega f(X_t/\Omega, c)$ where $\Omega$ is the cell volume. This results in
\begin{equation}\label{9}
d X_t = \Omega s f(X_t/\Omega, c) dt + \sqrt{\Omega s \text{diag}\{f(X_t/\Omega,
  c)\} s'} d W_t\; .
\end{equation}
Following \cite{Kampen2007}, we write the solution $X_t$ of the CLE as a
deterministic process $z_t$ plus a residual stochastic process,
\begin{equation}\label{10}
X_t = \Omega z_t + \sqrt{\Omega} M_t \;.
\end{equation}
Then, a Taylor expansion of the rate function around $z_t$ gives
\begin{equation}\label{12}
f(z_t + M_t/\sqrt{\Omega}, c) = f(z_t, c) + \frac{1}{\sqrt{\Omega}}F_t M_t +
O(\Omega^{-1}),
\end{equation}
where $F_t$ is the $v\times u$ Jacobian matrix with $(i, j)^{\text{th}}$ element $\partial f_{i}(z_t,c) / \partial z_{j,t}$ 
and $z_{j,t}$ is the $j$th component of $z_t$. Note that we suppress the dependence of $F_t$ on $z_t$ and $c$ for 
simplicity. Substituting (\ref{10}) and (\ref{12}) into equation~(\ref{9}) and collecting terms of 
$O(1)$ and $O(1/\sqrt{\Omega})$ give the ODE satisfied by $z_t$, and SDE satisfied by $M_t$ 
respectively, as
\begin{align}
dz_{t}&=s\,f(z_{t},c)dt, \label{13}\\
dM_{t}&=s\, F_t M_t dt + \sqrt{s\,\textrm{diag}\{f(z_t,c)\}s'}\,dW_t . \label{14}
\end{align}
Equations~(\ref{10}), (\ref{13}) and (\ref{14}) give the linear noise
approximation of the CLE and in turn, an approximation of the Markov jump process model.

For fixed or Gaussian initial conditions, the stochastic differential equation
in (\ref{14}) can be solved explicitly to give $(M_t\vert\, c) \sim N(m_t, V_t)$ 
where $m_t$ and $V_t$ satisfy the coupled deterministic system of ordinary differential equations
\begin{align}
\frac{d m_t}{dt} &= s F_t m_t,\label{16}\\
\frac{d V_t}{dt} &= V_t F_t' s' + s \text{diag}\{h(z_t, c)\} s' + s' F_t V_t \;. \label{17}
\end{align}
Hence, the approximating distribution of $X_t$ is as (\ref{gauss}) with
\begin{equation}\label{18}
\psi_t^l(c)=\Omega z_t + \sqrt{\Omega} m_t,\qquad \Sigma_t^l(c)=\Omega V_t \;.
\end{equation}
In situations where the ODE satisfied by $z_t$ is initialised with $z_0=x_0$ so 
that $m_0=0$, we see that $m_t=0$ for all $t$ and $\psi_t^l(c)=\Omega z_t$. 
Note further that $\Omega$ plays no role in the evolution equations (\ref{13}) and (\ref{17}). 
Therefore, in the examples in section 4, we assume a unit volume ($\Omega =1$) for simplicity.

\section{Diagnostic tools}


When model building, we usually want to investigate many different parameter
combinations. Similarly when inferring parameters, the data available is usually
limited and prior information on the plausible parameter values is sparse.
Therefore, parameter inference and model exploration usually follows a
combination of parameter scans, and/or exploring the parameter space using
efficient inference algorithms.

Since the parameter space to search will be large, it would be computationally
unfeasible to numerically assess the approximate simulator at all values. In
particular, since an approximate algorithm is being utilised, this implies that
simulating exact realisations may be computationally intensive. Thus the
parameter space must be explored efficiently.

One approach to explore the parameter space is random sampling, that is, we
sample uniformly in the parameter space. However, \cite{Mckay79a} showed that
Latin hypercube sampling (LHS) gave a significant improvement over simple random
sampling when exploring large spaces. \cite{Morris1995a} improved the original
LHS design with the maximin design, in which the distance between points in the
hypercube is maximised. Moreover, Latin hypercube sampling of the parameter space lends
  itself to an embarrassingly parallel mode of computation. Naturally, in scenarios 
that do not require a covering of the whole parameter space, other methods may be preferred. 
For example, if performing Bayesian inference via Markov chain Monte Carlo (MCMC) 
and focusing on regions of high posterior density, then we may choose to use the output 
of the MCMC scheme.    

\begin{figure}[t]
  \centering
  \includegraphics[width=0.4\textwidth]{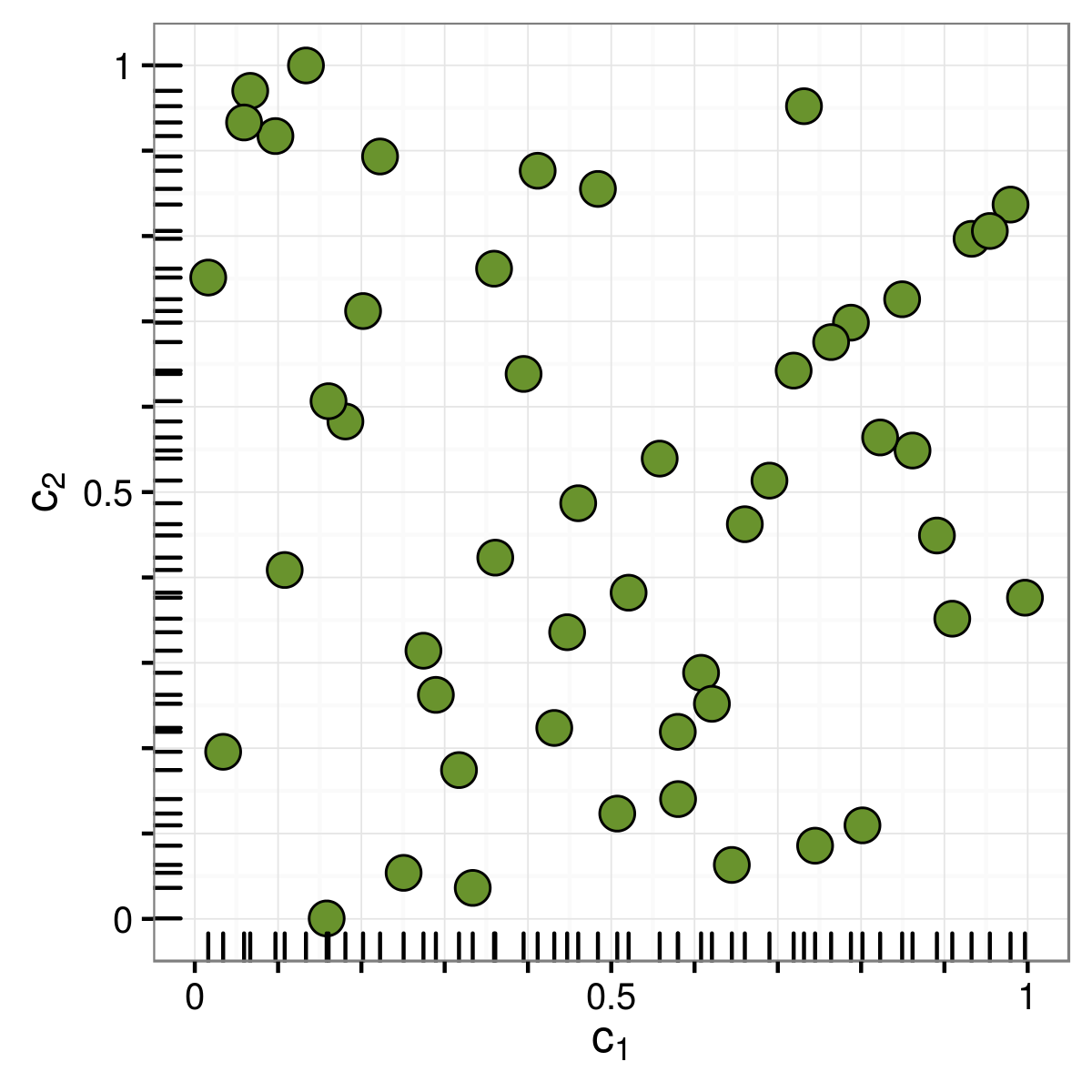}
  \caption{A two--dimensional Latin hypercube design, with $n_d = 50$
    points.}
  \label{F1}
\end{figure}

Figure \ref{F1} illustrates a two--dimensional design over parameters $(c_1,
c_2)$, with $n_d = 50$ points. We denote the $n_d$ points in the Latin hypercube as
\[
\mathbf{\gamma} = (\mathbf{\gamma}_1, \ldots, \mathbf{\gamma}_{n_d}) \;.
\]
Hence, each point $\mathbf{\gamma}_i$ is a length-$v$ (column) vector of parameter
values, that is $\mathbf{\gamma}_i=(c_{1,i},\ldots,c_{v,i})'$. A feature of these space filling designs, is that the marginal parameter
distributions have a uniform distribution, thereby giving good coverage in each dimension.

Our general strategy is to compare the moment closure/linear noise approximation
to a single realisation simulated exactly (using Algorithm~1) from the Markov 
jump process, at each of the $n_d$ points in the design. We refer to Algorithm~1 
as the exact simulator. For each design point $\mathbf{\gamma}_i$ let 
$\mathbf{x}^*(\mathbf{\gamma}_i) = (x_1^*(\mathbf{\gamma}_i),\ldots, x_u^*(\mathbf{\gamma}_i))'$ 
denote a single realisation from the exact simulator at a particular time point, 
with dependence on time, and the initial conditions used to produce the realisation, 
suppressed for ease of notation. In the following sections, we describe simple diagnostics 
that can be assessed by comparing the observed diagnostic at $\mathbf{x}^*(\mathbf{\gamma}_i)$ 
with the reference distribution of the diagnostic induced by the approximations described in (\ref{norm}).

\subsection{Individual prediction errors}\label{S3.2.1}

One way of assessing the accuracy of a Gaussian based approximation is to
calculate individual prediction errors. 
These are obtained by
calculating the difference between the exact simulator and the mean of the
linear noise (or two moment) approximation, that is
\begin{equation}\label{19}
e_{i,j} = x_j^*(\mathbf{\gamma}_i) - \psi_j(\mathbf{\gamma}_i)
\end{equation}
for each point $i=1,\ldots,n_d$ and species $j=1,\ldots, u$. Note that 
$\psi_j(\mathbf{\gamma}_i)$ denotes the $j$th component of the mean in (\ref{gauss}) 
after omitting dependency on time $t$. Plainly, a more appropriate quantity to work with is the standardised prediction
error
\begin{equation}\label{20}
  e_{i,j}^* = \frac{x_j^*(\mathbf{\gamma}_i) - \psi_j(\mathbf{\gamma}_i)}{\sqrt{\Sigma_{jj}(\mathbf{\gamma}_i)}} \;.
\end{equation}
If $\mathbf{x}^*(\mathbf{\gamma}_i)$ is replaced with a draw from 
either the two moment or linear noise approximation, then the standard prediction 
errors can be seen as draws from a standard normal distribution. Hence, 
large standardised individual errors, with absolute values larger than say two, 
indicate a potential discrepancy between the exact
and approximate simulators. Of course, single, isolated values are possible, and
so further investigation can be performed by obtaining more simulator runs in the
parameter vicinity. 

Since the reference distribution of the standardised prediction errors is normal, 
we can use other standard techniques for assessing the modelling assumptions 
that underpin both approximate simulators. For example, 
quantile-quantile (q-q) plots provide a natural graphical diagnostic for assessing 
normality, with a reasonable fit indicated by points close to a 45-degree line through the 
origin. We may expect the output of the exact simulator to be heavier tailed than a Gaussian, 
in which case points in the q-q plot will cluster around a line with a slope greater than one. 
Plotting errors against parameter values may also be useful in identifying regions of parameter 
space that exhibit large discrepancies.

We note that at each point on the Latin hypercube, it is possible to draw $n_{ex}$ realisations from
an exact simulator (giving a total of $N = n_d \times n_{ex}$ exact
simulations), and use a formal hypothesis test in the spirit of
\cite{Jenkinson2013a}. However, there are a number of potential drawbacks with
this approach. First, the computational cost may be prohibitively large. For a
fixed computational budget of $N$ simulations, either $n_{ex}$ would be
prohibitively small, which would adversely affect the power of the test, or we
would reduce $n_d$ and not explore the parameter space. Second, for both the LNA
and 2MA schemes, if the model contains a second order reaction, we can prove
analytically that the mean and variance are not equal to the true value, so a
hypothesis test is not needed. Furthermore, any non-significant test must be
spurious. Third, the normal assumption is also clearly incorrect since the state
space is discrete. Therefore, we focus on an assessment of whether the
approximation is ``good enough'' over a large parameter space.

\subsection{Interval diagnostic}

Another straightforward method for assessing fit is to construct a $100\alpha$\%
confidence interval for $x_j^*(\mathbf{\gamma}_i)$ using the 
mean $\psi(\cdot)$ and variance $\Sigma(\cdot)$ associated with the approximate simulator under assessment.
We denote a particular confidence interval at design point $i$, for species $j$,
as $CI_{i,j} (\alpha)$. The proportion of simulated values that land within the
confidence region is given by
\begin{equation}\label{7b}
  D_j^{CI} = \frac{1}{n_d} \sum_{i=1}^{n_d} 1[x_j^*(\mathbf{\gamma}_i)
\in CI_{i,j}(\alpha)],
\end{equation}
where $1[\cdot]$ is the indicator function. We can assess fit as the value of
$D_{j}^{CI}$ should be approximately equal to $\alpha$. Additionally, plotting
the confidence regions against parameter values can highlight any particular
systematic deviations.

\subsection{LNA vs 2MA}\label{lnaVSmc}

Recently \cite{Grima2012} explored the link between the two moment and linear
noise approximations. Essentially, the two approximations are very similar,
except that the mean equations in the LNA do not depend on the covariances. This
would suggest that if the two approximations gave appreciably different
estimates for the first two moments, further investigation is required.

We define the standardised difference between the two approximations as
\begin{equation}\label{22}
  D_{j}^{LM}(\mathbf{\gamma}_i) = \frac{\psi_j^l(\mathbf{\gamma}_i) - \psi_j^m(\mathbf{\gamma}_i)}{\sqrt{\Sigma_{jj}^l(\mathbf{\gamma}_i)}}\;.
\end{equation}
Note again that for notational convenience, the time subscript $t$ has been omitted from
the expression. Large differences of $D_{j}^{LM}$ should be carefully
investigated. This diagnostic measure has the advantage of avoiding (possibly
expensive) exact simulation. However, when the two approximations give similar
results, it does not necessarily follow that both approximations are correct. For 
example, \cite{Schnoerr2014} highlighted an oscillating system where the LNA
and 2MA schemes were in agreement, but were significantly different from the
solution to the underlying chemical master equation.

\section{Examples}

Here, we demonstrate the diagnostic tools in three examples. Diagnostics based
on the linear noise approximation are constructed for two reaction networks that
are known to exhibit interesting non-linear dynamics. In the final example, we
consider the prokaryotic auto regulatory gene network analysed by
\cite{Golightly:2008} and \cite{Milner2012}. We focus on the moment closure
approximation and construct diagnostics to assess approximate simulator fit both
\emph{a priori} and \emph{a posteriori}. Interactive versions of all graphics
can be found at
\begin{center}
\url{https://bookdown.org/csgillespie/diagnostics/}
\end{center}

\subsection{Schl\"ogl system}\label{schlogl}

The Schl\"ogl model is a well known test system that exhibits bi-modal and
non-linear characteristics at certain parameter combinations. The system
contains four reactions

\begin{center}
  \begin{tabular}{@{} >{$}r<{$\!\!\!} 
      >{$}r<{$} >{$}l<{$} >{$}l<{$} 
      >{$\qquad}r<{$\!\!\!} 
      >{$}r<{$} >{$}l<{$} >{$}l<{$}
      @{}}
    R_1\text{:} 
    &  2X_1 + X_2
    & \xrightarrow{\phantom{a}c_{1}\phantom{a}}
    &  3X_1
    & R_2\text{:}   
    & 3X_1
    & \xrightarrow{\phantom{a}c_{2}\phantom{a}} 
    & 2X_1 + X_2
    \\
    R_3\text{:} 
    & X_3 
    & \xrightarrow{\phantom{a}c_{3}\phantom{a}} 
    & X_1
    & R_4\text{:} 
    & X_1
    &\xrightarrow{\phantom{a}c_{4}\phantom{a}} 
    & X_3 \\
  \end{tabular}
\end{center}
describing the evolution of three chemical species, $X_1$, $X_2$, and $X_3$ and
assumes mass action kinetics. In this example, we concentrate on species $X_1$.
Where the distribution of $X_1$ is bi-modal, the linear noise approximation
would clearly be inappropriate. However for large models, it isn't necessary
clear if (or where) a system would have bi-modal regions. Hence, the purpose of
this example is to illustrate how problematic regions may be detected.

\begin{figure}[!t]
  \centering
  \includegraphics[width=0.4\textwidth]{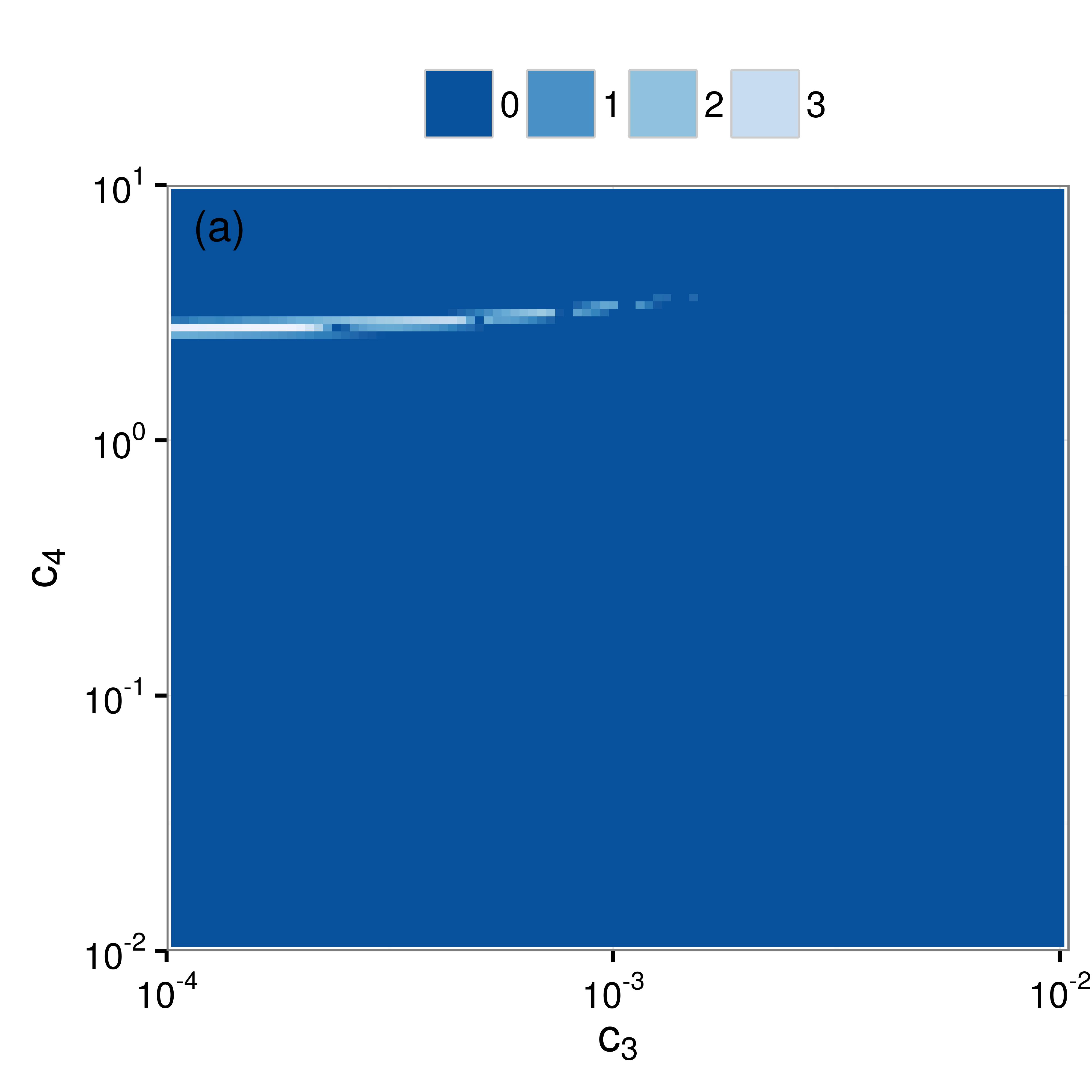}%
  \includegraphics[width=0.4\textwidth]{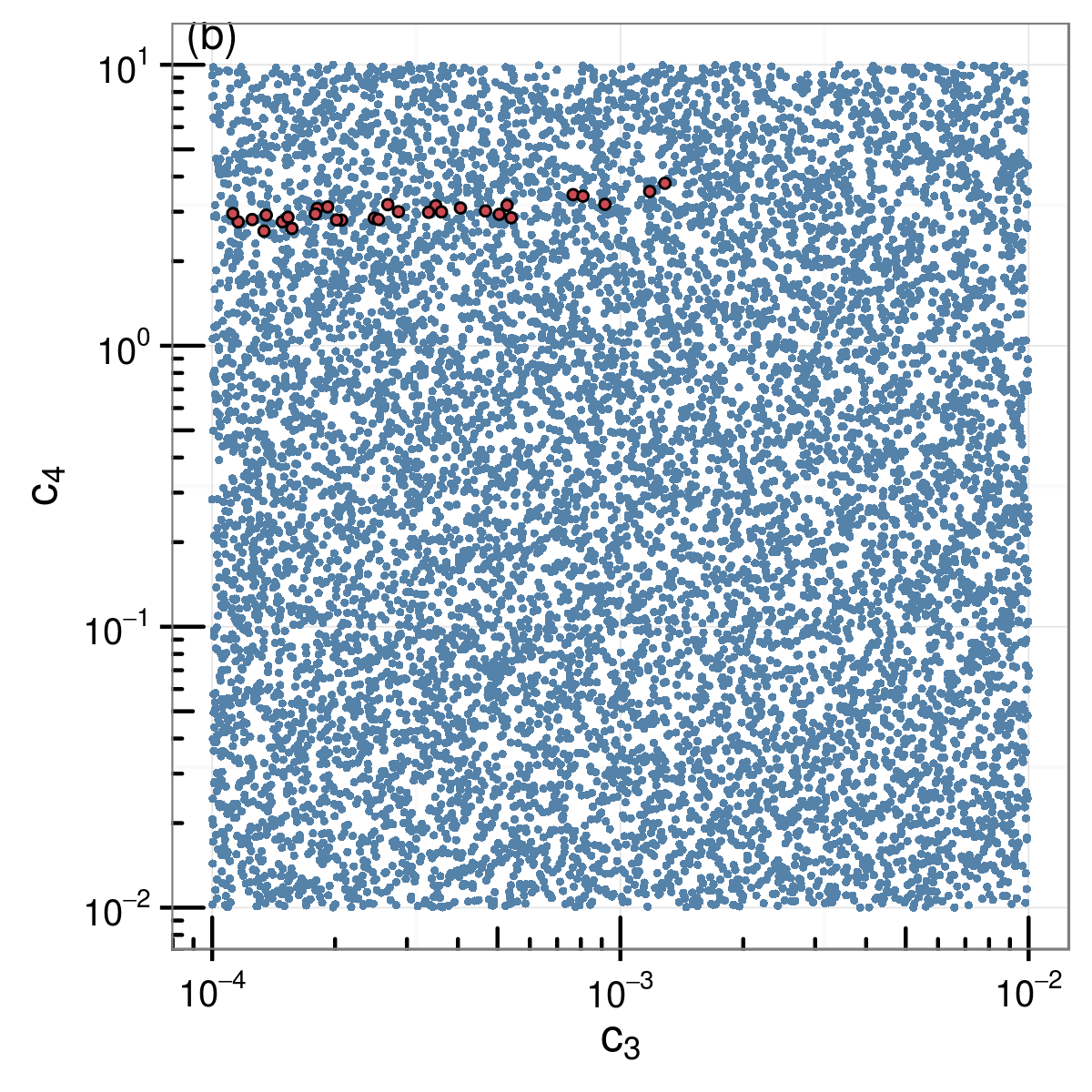}\\
  \includegraphics[width=0.4\textwidth]{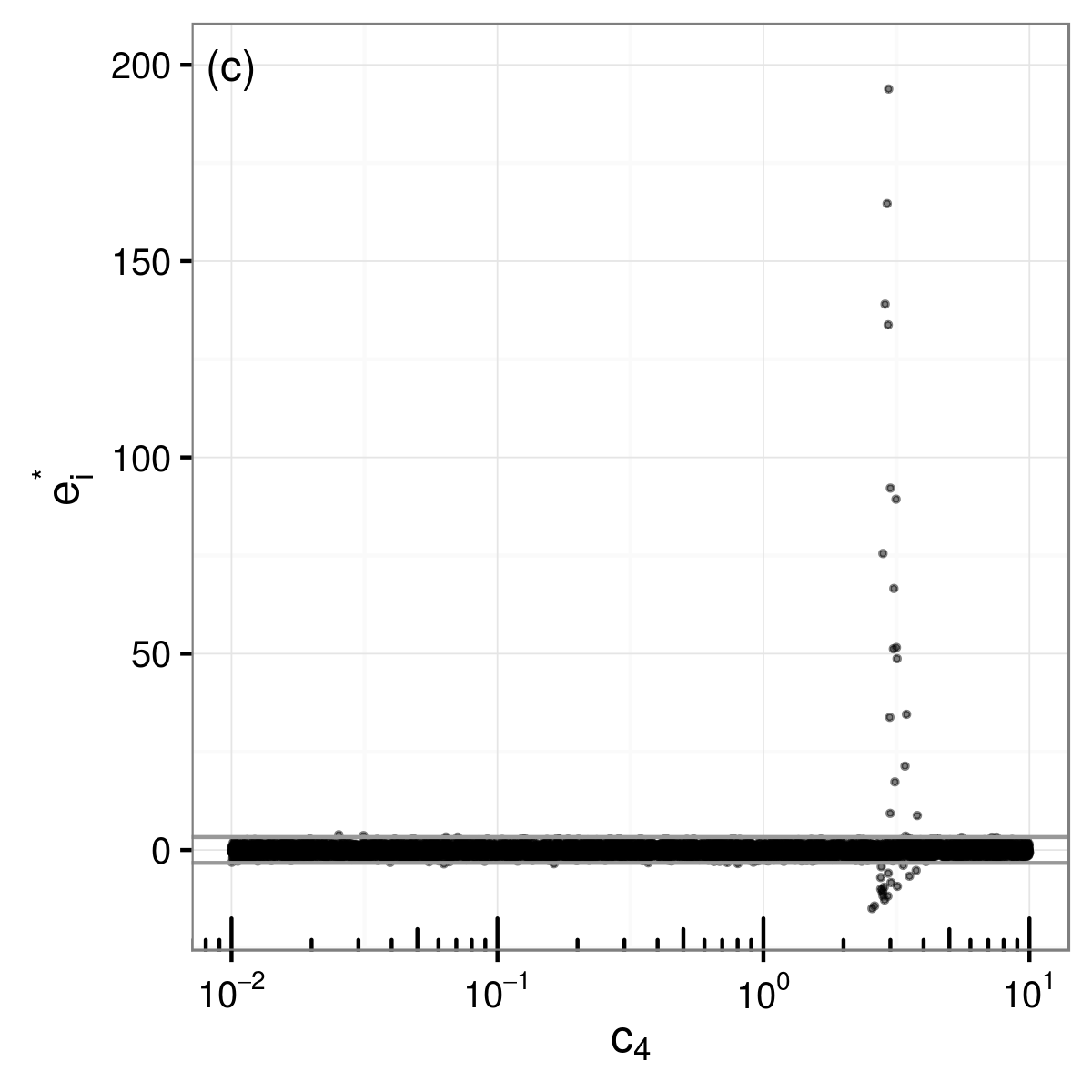}%
  \includegraphics[width=0.4\textwidth]{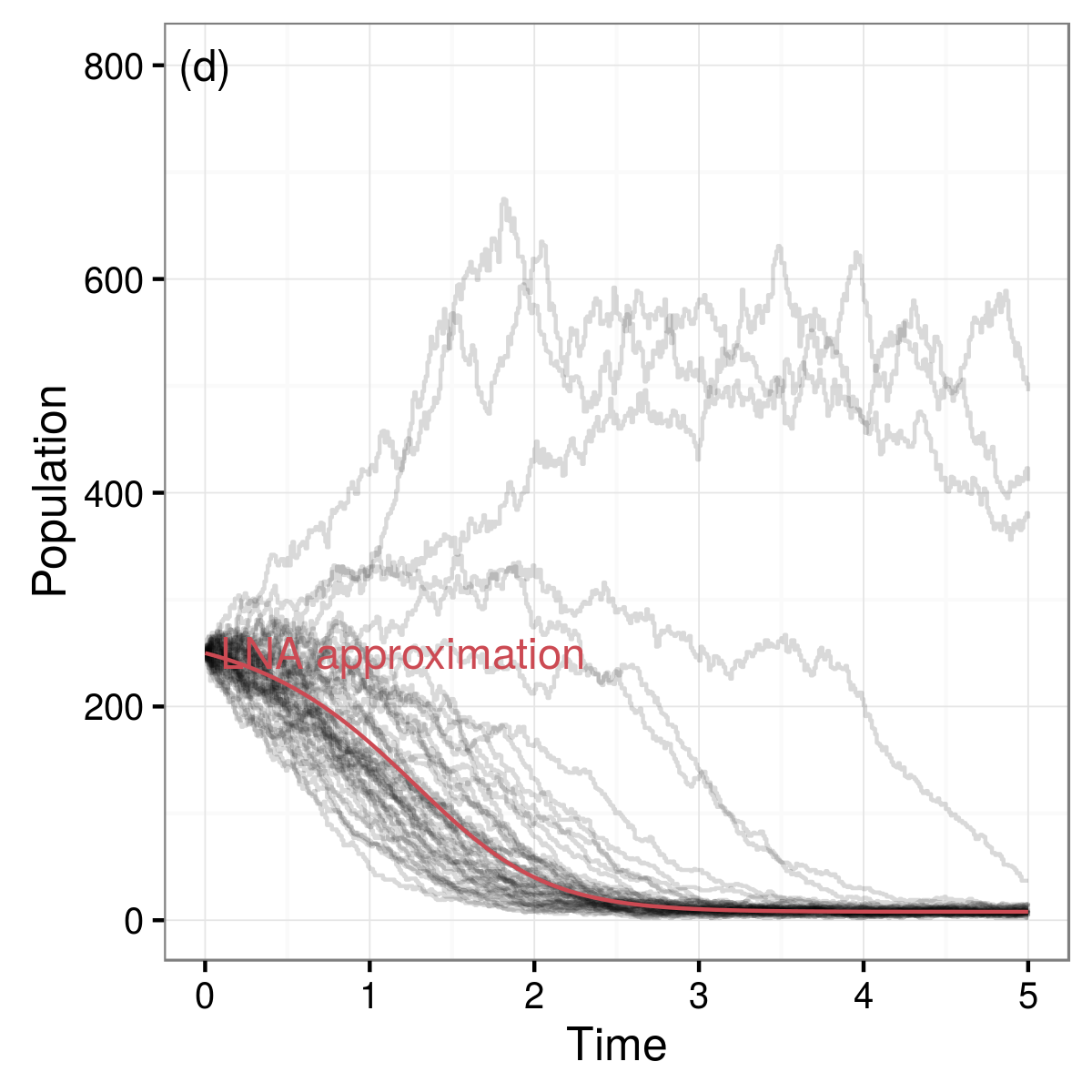}%
  \caption{Model diagnostics for the Schl\"ogl system. (a) Image plot
    highlighting the bi-modality region for the Schl\"ogl system. (b) Latin
    hypercube design (on the $\log_{10}$ scale), with $n_d=10,000$. The largest
    $30$ errors are shown in red. (c) Standardised prediction errors,
    with $95$\% and $99.9$\% regions indicated by grey lines. (d) Fifty stochastic simulations, where $c = (3\times
    10^{-7}, 10^{-4}, 0.00011, 2.955)$ (the parameter values associated with the
    largest prediction error). The mean of the linear noise approximation is shown in
    red.}\label{F2}
\end{figure}

The parameters $(c_1, c_2)'$ were fixed at $(3\times 10^{-7}, 10^{-4})'$ and
the initial conditions assumed constant at
\[
x_0 = (250, 10^5, 2\times 10^5)' \;.
\]
Suppose that interest lies in the accuracy of the linear noise approximation at 
time-point $t=5$. Further, consider a parameter space defined by the regions 
$c_3 =(10^{-4}, 10^{-2})$ and $c_4 = (10^{-2}, 10)$ (on the $\log_{10}$
scale). Figure~\ref{F2}a shows the region in parameter space that leads to 
a bi-modal distribution of $X_1$. The plot was obtained by finely discretising the 
parameter space (to give a $1000\times 1000$ grid) and calculating the absolute 
prediction error (on the $\log_{10}$ scale) at each parameter value. For systems of 
realistic size and complexity, this approach will be computationally prohibitive. 

We therefore generated a Latin hypercube with $n_d=10,000$ points. The two dimensional design
space is given in Figure \ref{F2}a. The standardised prediction errors plotted against parameter $c_4$ are shown in
Figure \ref{F2}b. The locally smoothed mean value (shown in blue), is close to
zero. However there are several large errors, in particular, $e_{8684}^* \simeq
194$. This large error was further investigated using fifty realisations from the
exact simulator with the parameter values set at $\mathbf{\gamma}_{8684}$ (see
Figure \ref{F2}d). The LNA mean solution is also shown in red. It is clear that
at this particular choice of parameter values, the Schl\"ogl system has a
bi-modal distribution and the LNA is inappropriate in this region of parameter
space. Therefore, with relatively few design points, we are able to detect regions 
of parameter space that lead to significant discrepancies between the exact and approximate 
simulators. Naturally, care must be taken in the choice of $n_d$ and this will typically 
be dictated by computational budget. We find that for this example, reducing $n_d$ to 1000 
results in only a single value in the Latin hypercube design with an absolute prediction 
error greater than 2. 


As discussed in section \ref{S3.2.1}, rather than generate a single exact simulation at
each of the $n_d$ points, we could simulate $N$ times, where $N = n_{d}
\times n_{ex}$, to give $n_{ex}$ replicates at each of the design points, allowing comparison of the 
simulator output via a formal hypothesis test. This is similar to the example in \cite{Jenkinson2013a}, where
the authors set $n_{ex} = 350$. We note that a computational budget allowing $N=10,000$ 
would result in only 30 points in the hypercube being assessed. It is highly unlikely, in this case, 
that the region of bi-modality would be detected.

\subsection{Lotka-Volterra model}\label{LV}

\begin{figure}[t]
  \centering 
  \includegraphics[width=0.4\textwidth]{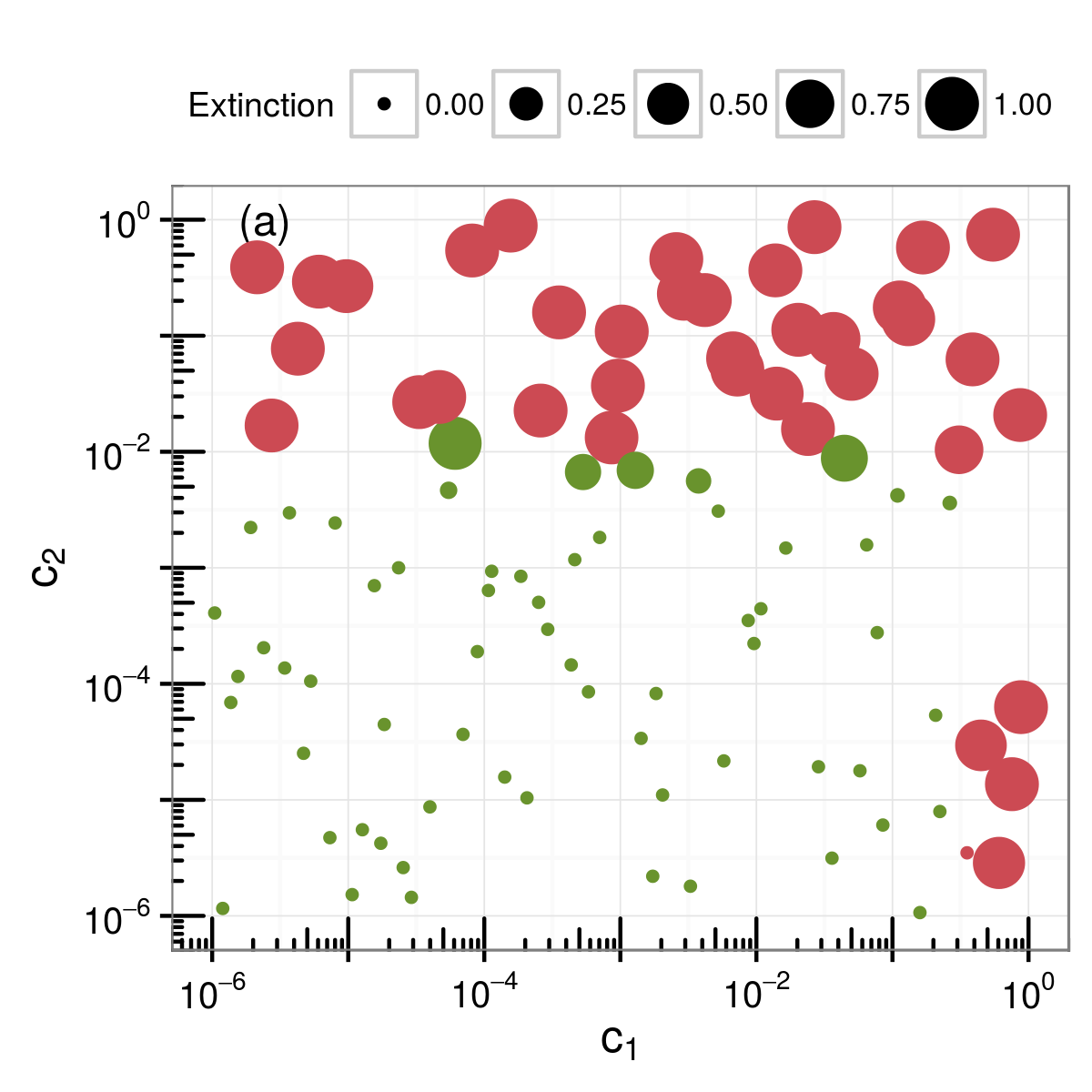}%
  \includegraphics[width=0.4\textwidth]{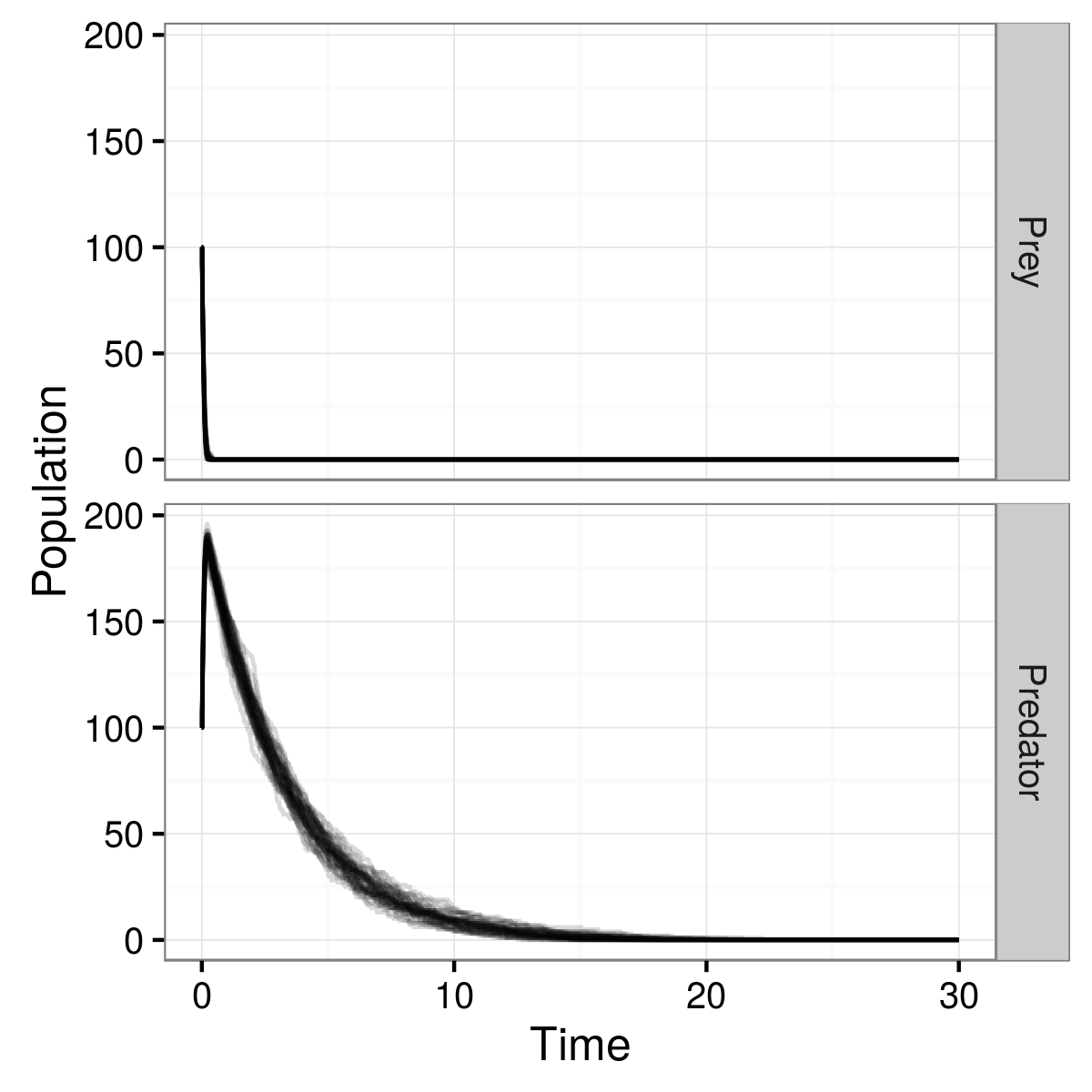} 
  \caption{Lotka-Volterra predator prey system. All simulations used initial
    conditions $x_0 = (100,100)'$. (a) Latin hypercube design (with
    $n_d=100$) on the $\log_{10}$ scale. At each point on the hypercube the
    $D^{LM}$ diagnostic was calculated. Values where $\vert D_{i, 2}^{LM}\vert > 5$ are
    shown as red circles. The radius of each circle is proportional to the probability 
 of prey extinction by time 30. (b) Fifty stochastic simulations with parameter
    values $c = (10^{-4}, 0.1, 0.3)'$.}\label{F3}
\end{figure}

The predator prey system developed by \cite{lotka:25} and \cite{volterra:26},
describes the time evolution of two species, $X_1$ and $X_2$. This system has two species and three reactions
\begin{center}
  \begin{tabular}{@{} >{$}r<{$\!\!\!} 
      >{$}r<{$} >{$}l<{$} >{$}l<{$} 
      >{$\qquad}r<{$\!\!\!} 
      >{$}r<{$} >{$}l<{$} >{$}l<{$}
      @{}}
    R_1\text{:} &
    X_{1}  & 
    \xrightarrow{\phantom{a}c_{1}\phantom{a}} &
    2X_{1}  &
    R_{2}\text{:} &
    X_{1} + X_{2} &
    \xrightarrow{\phantom{a}c_{2}\phantom{a}} &
    2 X_{2}  \\
    R_3\text{:} &
    X_{2}  & 
    \xrightarrow{\phantom{a}c_{3}\phantom{a}} &
    \emptyset \;. \\
  \end{tabular}
\end{center}
Although relatively simple, this system exhibits interesting auto regulatory 
behaviour and has been used numerous times to test inference algorithms;
see, for example, \cite{Boys:2008,Opper:2008,White2013}. In particular, the
linear noise and two moment approximations have been used for parameter
inference \citep{Milner2012,Golightly2014}. 

To assess the linear noise approximation, we generated $n_d = 100$ points from a
two-dimensional Latin hypercube, over the regions $c_1 = (10^{-6}, 10^0)$ and
$c_2 = (10^{-6}, 10^0)$ on the $\log_{10}$ scale. These regions correspond to an
inference situation where we are using vague priors. We set $c_3 = 0.3$ and used
initial conditions $x_0 = (100,100)'$ with a maximum simulation time of $t=30$.
Figure \ref{F3}a shows the Latin hypercube design. The diagnostic of Section~\ref{lnaVSmc} 
was computed at each design point. Values where $\vert D_{i, 2}^{LM}\vert > 5$ are
    shown as red circles in Figure~\ref{F3}. The radius of each circle is proportional to the probability 
 of prey extinction by time 30. It is clear that for large
values of $c_1$ or $c_2$, the LNA and 2MA approximations disagree. Moreover, we see that 
these points coincide with a high probability of prey extinction by time 30 (see also 
Figure \ref{F3}b, showing fifty realisations from the exact simulator at a typical discrepant 
parameter value). This result is perhaps unsurprising given the time-course behaviour of the Markov 
jump process representation of the Lotka-Volterra system. The system eventually reaches 
one of two states: if $X_1$ dies out then the system will run to $(0,0)$ (reactions 1 and 2 will never again occur). 
If $X_2$ dies out the system will go towards $(\infty, 0)$ (reactions 2 and 3 will never again occur). The LNA 
fails to capture this behaviour. For example, the LNA mean is a perfectly repeating oscillation, carrying 
on indefinitely. As expected, increasing $t$ leads to a higher proportion of the parameter space 
with significantly large prediction errors (results not reported).

\subsection{Prokaryotic auto regulatory gene network}\label{prok}

A more realistic example is the prokaryotic auto regulation system. This larger
model contains six species and twelve reactions. In this network a protein
$I$ coded for by a gene $i$ represses its own transcription and also the
transcription of another gene $g$ by binding to a regulatory region upstream
of the gene. This is described by the reactions
\begin{align*}
  R_1\!:~& I +i \rightarrow I\cdot i, &
  R_2\!:~& I\cdot i \rightarrow I+i, \\
  R_3\!:~& I+g \rightarrow I\cdot g, & 
  R_4\!:~& I\cdot g \rightarrow I + g .
\end{align*}
The transcription of $i$ and $g$ and the translation of mRNA
$r_i$ and $r_g$ are represented by
\begin{align*}
  R_5\!:~&i\rightarrow i + r_i, &
  R_6\!:~&r_i\rightarrow r_i + I,\\
  R_7\!:~&g\rightarrow g + r_g , & 
  R_8\!:~&r_g\rightarrow r_g + G.
\end{align*}
We also have mRNA degradation
\begin{align*}
  R_9\!:~& r_i \rightarrow \emptyset, & R_{10}\!:~& r_g\rightarrow \emptyset,
\end{align*}
and protein degradation
\begin{align*}
  R_{11}\!:~& I \rightarrow \emptyset, & R_{12}\!:~& G \rightarrow \emptyset .
\end{align*}
Each reaction $i$ has a stochastic rate constant $c_i$. There are two
conservation laws in the model
\begin{align*}
  I & \cdot i + i = K_1, & I & \cdot
  g + g =K_2,
\end{align*}
where $K_1$ and $K_2$ are conservation constants. If $K_1$ and $K_2$ are known,
then we can simplify the model using the conservation laws to remove $I
\cdot i$ and $I \cdot g$. This simplification reduces
the model to six species
\[
x = (I, G, i, g,r_i,r_g)' \;.
\]
The reaction hazards for $R_1$ and $R_2$ are $h_1(x, c_1)=c_1 I i$ and
$h_2(x,c_2)= c_2 I\cdot i = c_2 (K_1 -i)$ respectively. Hazards for $R_3$ and
$R_4$ are calculated similarly. The remaining hazards are for first order reactions.

This model has been used to test parameter inference schemes by
\cite{Golightly:2008} and \cite{Milner2012}. In this example, we will explore the moment
closure approach used by \cite{Milner2012}.

\begin{figure}[!t]
  \centering
  \includegraphics[width=0.8\textwidth]{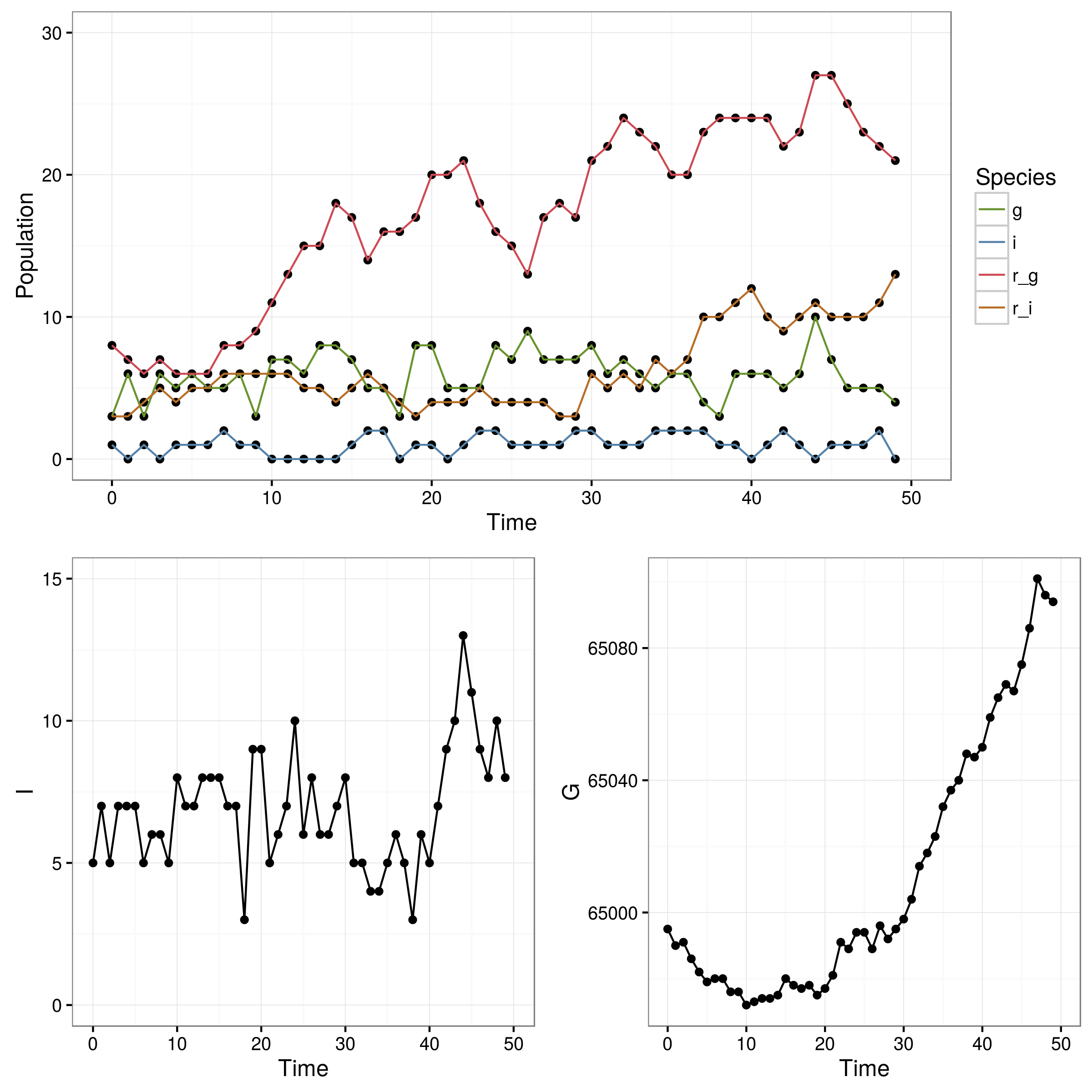}
  \caption{A single stochastic realisation from the prokaryotic auto regulatory
    gene network. This realisation was observed at times $t=0, 1, \ldots, 49$.}\label{F4}
\end{figure}
 
The stochastic realisation that \cite{Milner2012} based their parameter inference
on is given in Figure \ref{F4}. Many of the chemical species have population
sizes less than twenty. However, the population of species $G$ has a
population greater than $65,000$. Hence, exact simulations based on parameter 
values consistent with the data in Figure \ref{F4} are computationally expensive. This 
prohibits the use of inference algorithms based on exact simulation.

\subsubsection{Inference set-up}

We use a data set of fifty observations at (unit) discrete time points of the
simulated process (see Figure \ref{F4} for the trace of the realisation). The
true parameter values for $(c_1, c_2, \ldots, c_{12})$ that produced the data
set were $(0.08, 0.82, 0.09, 0.9, 0.75,$ $0.05, 0.35, 0.5, 0.1, 0.1, 0.05,
0.0001)$. It is worth noting that gene $i$ has at most two copies and
only takes values 0, 1 or 2.

Only vague prior knowledge was assumed about parameter values, with Uniform
$U(-5, 1)$ priors for each $\log(c_i)$ for $i=1, \ldots, 12$ and $U(-12, -6)$ on
$\log(c_{12})$. The values of $K_1$ and $K_2$ were assumed known and set at two
and ten respectively.

\subsubsection{Prior investigation}
\begin{figure}[!t]
  \centering
  \includegraphics[width=0.85\textwidth]{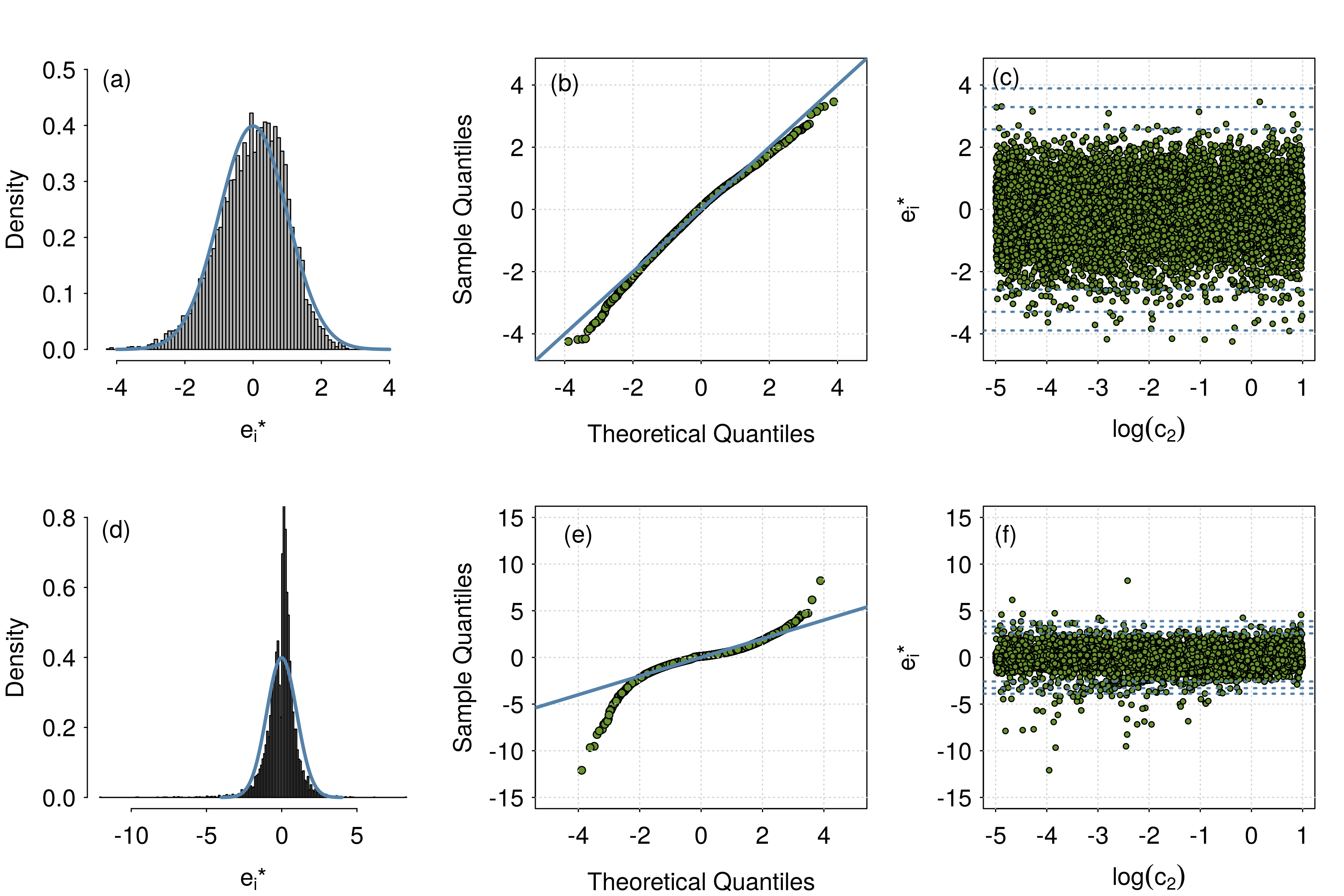}
  \caption{Predictive error plots for \texttt{I} (top row) and \texttt{i}
    (bottom row) based on the prior distribution. A total of $n_d=10,000$ points
    were sampled from a thirteen dimensional Latin hypercube (twelve parameters
    plus the time dimensional). The blue lines on figures (a), (b), (d) and (e)
    represent the standard normal distribution. The dashed lines in figures (c)
    and (f) represent the 99\%, 99.9\% and 99.99\% regions in the standard
    normal distribution.}\label{F5}
\end{figure}

An $n_d = 10,000$ point twelve dimensional Latin hypercube was created on the log
space over the parameter prior regions. At each point on the hypercube, a time
point from the realisation in Figure \ref{F4} was selected to initialise the
exact simulator and the moment closure approximation. Each simulator was then
run for a single time point and the standardised prediction error was
calculated.

Figure \ref{F5} (a)--(c) gives the diagnostic plots for species \texttt{I}.
Although the population levels of $I$ are relatively small, the
population size varies between 3 and 13, the associated diagnostic plots still
look reasonable. The diagnostic plots for species $i$ are given in Figure \ref{F5}
(d)--(f). This species only takes values 0, 1, and 2. As would be expected, the
diagnostic plots show clear deviations from the normality assumptions. In
particular, when $c_2 << 0.1$, we obtain a number of very large standardised
prediction errors. As with the Schl\"ogl system, it would be advisable to
investigate these problematic points more carefully. We note that for the data set 
in Figure~\ref{F4}, 
the marginal posterior density for $c_2$ has negligible mass in this region of 
parameter space (the true value of $c_2$ is 0.82).

\subsubsection{Posterior investigation}

A further investigation of the appropriateness of the moment approximation can
be made \emph{a posteriori}. Since the parameters in the posterior
distribution were in some cases highly correlated, we sampled $n_d = 10,000$
points from this posterior. 

\begin{figure}[!t]
  \centering
  \includegraphics[width=0.85\textwidth]{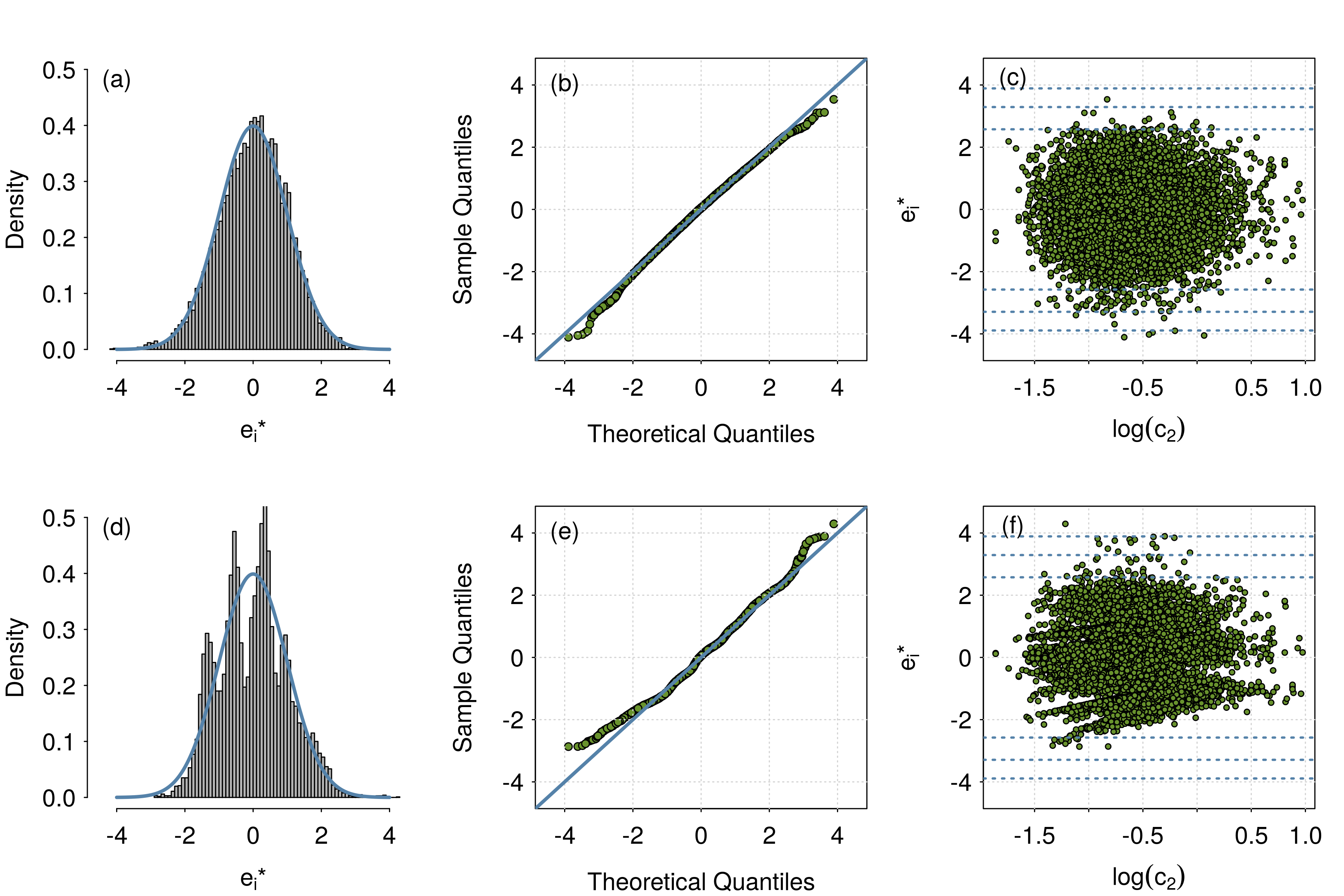}
  \caption{Predictive error plots for \texttt{I} (top row) and \texttt{i}
    (bottom row), based on the posterior distribution. A total of $n_d=10,000$
    points were sampled from the posterior distribution. The blue lines on
    figures (a), (b), (d) and (e) represent the standard normal distribution.
    The dashed lines in figures (c) and (f) represent the 99\%, 99.9\% and
    99.99\% regions in the standard normal distribution.}\label{F6}
\end{figure}

Again, the diagnostic plots for species $I$ (Figure \ref{F6} (a)--(c))
suggest that the normality assumption and the accuracy of the mean and variances of
the moment closure approximation appear reasonable. The diagnostic plots for low level species $i$ have substantially
improved (see Figure \ref{F6} (d)--(e)), although we observe extreme
standardised errors in Figure \ref{F6} (f). Of course, since $i$ can
only take values 0, 1, and 2, the prediction errors are not normally distributed
(see Figure \ref{F6} (d)). Although the moment closure approach fails 
to adequately match the Markov jump process in all regions of parameter space \emph{a priori}, 
in regions of high posterior density, it does appear to provide a satisfactory alternative.

\section{Discussion}

Analysing stochastic kinetic models of realistic size and 
complexity is a challenging problem. For example, whilst it is possible, in principle, 
to perform exact (simulation-based) inference for the Markov jump process (MJP) 
representation \citep{Boys:2008,Golightly:2011,Owen:2015}, existing approaches 
are computationally intensive and have ostensibly focused on toy examples 
with relatively few numbers of species and reactions. Replacing the exact 
MJP simulator with a cheap approximation and using this for model exploration/inference 
is an appealing alternative approach. Gaussian approximations that ignore discreteness 
but not stochasticity, such as the linear noise approximation (LNA) and moment closure 
approaches considered here, are particularly attractive due to their tractability. 
While this assumption can make inference easier, it is
essential to assess the appropriateness of the Gaussian approximation. It is apparent 
from the literature that such an assessment rarely takes place.

In this paper we have presented a general, easy-to-use, framework that allows modellers to
determine whether a given Gaussian approximation is suitable for their model. Following the approach 
of \cite{Bastos2009c}, we have examined simple numerical diagnostics, by constructing appropriate 
functions of the exact simulator output. Comparing observed values of the diagnostic (for a particular 
parameter value) to the distribution induced by the approximation gives an indication of whether 
or not the approximation can adequately represent the MJP. By using
efficient space filling designs to explore the parameter space, we can assess an
approximate simulator across a large region. In particular, since each point in
the Latin hypercube design can be simulated independently, we can use
cloud computing to explore vast regions of the parameter space. 

We applied our 
approach to three examples in which the underlying Markov jump process exhibits 
interesting non-linear dynamics. For the Schl\"ogl system (Section~\ref{schlogl}), our approach was able 
to detect a region of bi-modality using relatively few design points. In the 
Lotka-Volterra example (Section~\ref{LV}), a comparison of the linear noise approximation 
and moment closure approach was able to identify regions of the parameter space that lead 
to prey extinction. Finally, for the prokaryotic auto regulatory gene network (Section~\ref{prok}), 
we considered the synthetic data set of \cite{Milner2012} and 
compared the moment closure approach with the MJP over parameter regions determined 
both \emph{a priori} and \emph{a posteriori}. We found that in regions of high 
posterior density, the approximation does appear to provide a satisfactory 
alternative to the MJP, despite the inherent discreteness of the observed data. 
 
\section*{Computing details}

All simulations were performed on a machine with 16GB of RAM and with an Intel
quad-core CPU. The operating system used was Ubuntu 12.04. Simulations for the
Lotka-Volterra model and Schl\"ogl system were performed using R (version
3.3.1), via the \texttt{issb} package (version 0.13.3)
(\cite{RCoreTeam2013,Golightly2013b}. The Latin hypercube was generated using
the \texttt{lhs} package (version 0.13) \citep{Carnell2012}. The graphics were
created using the \texttt{ggplot2} R package (version 2.1.0)
\citep{Wickham2009}. The Prokaryotic auto regulatory gene network code used a
combination of C (from the \cite{Milner2012} paper) and R code. 
\bigskip

\noindent\textbf{Acknowledgements:} We thank the three anonymous referees for 
constructive comments that have improved the paper.

\bibliography{refs}

\end{document}